\documentclass[conference,a4paper]{IEEEtran}
\usepackage[T1]{fontenc}
\usepackage{newunicodechar}
\newunicodechar{–}{\textendash}
\IEEEoverridecommandlockouts
\usepackage{cite}
\usepackage{amsmath,amssymb,amsfonts}
\usepackage{algorithm}
\usepackage{algpseudocode}
\usepackage{graphicx}
\usepackage{textcomp}
\usepackage{xcolor}
\usepackage{url}
\usepackage{multirow}
\usepackage[a4paper, total={184mm,239mm}]{geometry}
\usepackage{fancyhdr}
\fancypagestyle{icecsheader}{
    \fancyhf{}
    \fancyhead[C]{%
        \fontsize{8}{9}\selectfont
        2026 IEEE 33rd International Conference on Electronics, Circuits and Systems (ICECS)}

}
\fancypagestyle{icecsfirstpage}{
    \fancyhf{}
    \fancyhead[C]{%
        \fontsize{8}{9}\selectfont
        2026 IEEE 33rd International Conference on Electronics, Circuits and Systems (ICECS)}
    \fancyfoot[L]{%
        \fontsize{8}{9}\selectfont
        979-8-3195-1905-4/26/\$31.00~\textcopyright{}~2026 IEEE}

}
\usepackage{makecell}
\usepackage{booktabs}

\def\BibTeX{{\rm B\kern-.05em{\sc i\kern-.025em b}\kern-.08em
    T\kern-.1667em\lower.7ex\hbox{E}\kern-.125emX}}
\usepackage{todonotes}

  \renewcommand{\baselinestretch}{0.955}
\begin{document}

\title{FINNAS: FINN-Guided Hardware-Aware NAS and Pruning for FPGA Jet Substructure Classification}


\author{
Eva Chauffour,
Changhong Li,
Georgios Floros,
Shreejith Shanker \\ 
Reconfigurable Computing Systems Lab, Electronic \& Electrical Engineering\\
Trinity College Dublin, Ireland\\
Email: \{chauffoe, lic9, florosg, shreejith.shanker\}@tcd.ie
}
\maketitle
\IEEEpubid{%
\makebox[\columnwidth][l]{%
\fontsize{8}{9}\selectfont
979-8-3195-1905-4/26/\$31.00~\textcopyright{}~2026 IEEE}
\hspace{\columnsep}
\makebox[\columnwidth]{}}
\thispagestyle{icecsfirstpage}


\begin{abstract}
FPGAs are well suited to deploying quantised neural networks (QNNs)
under strict accuracy, latency, and resource constraints; however, identifying efficient model-accelerator combinations commonly requires extensive manual design-space exploration and repeated hardware synthesis. This paper presents FINNAS, a FINN-guided hardware-aware evolutionary neural architecture search framework. FINNAS jointly searches quantised MLP depth, width, and global precision settings, and ranks candidates using proxy validation accuracy together with FINN-estimated LUT usage and latency under a fully parallel mapping. Selected finalists are fully retrained, subjected to post-search unstructured pruning, and validated using RTL simulation and Vivado out-of-context synthesis. On the CERNBox jet substructure classification task, the searched implementations expose competitive accuracy-resource trade-offs. Compared with a manually optimised dense FINN accelerator, a compact FINNAS design improves accuracy from 73.78\% to 74.36\%, while reducing LUT usage by \(8.5\times\) and RTL-simulation latency by \(1.77\times\). Unstructured pruning further provides consistent LUT and FF reductions across the fully parallel finalists.
\end{abstract}

\section{Introduction}\label{introduction}


The convergence of deep learning and edge computing has created a strong demand for low-latency neural network inference on specialised hardware. 
In latency-critical scientific applications, inference must often be completed within submicroseconds while satisfying resource constraints.
High-energy physics experiments are among the most representative scenarios~\cite{duarte2018fast}, where jet substructure classification (JSC) is a widely used benchmark for real-time FPGA inference. 
Specifically, the HLF CERNBox JSC formulation represents each event using 16 high-level input features and is commonly evaluated using compact fully connected classifiers.


Several FPGA-oriented implementations have been developed for JSC, including LogicNets~\cite{umuroglu2020logicnets}, PolyLUT~\cite{andronic2025polylut}, NeuraLUT~\cite{andronic2024neuralut}, NeuraLUT-Assemble~\cite{andronic2025neuralutassemble}, and AmigoLUT~\cite{weng2025amigolut}, which map neurons directly to logical lookup-table operations to exploit FPGA primitives.
These approaches can achieve high classification accuracy with very limited resource utilisation while meeting the real-time requirements of the task.
However, these accelerators typically rely on specialised mappings and relatively complex implementation processes.


More general FPGA QNN deployment frameworks like FINN~\cite{blott2018finn} provide an end-to-end solution for neural network acceleration on FPGAs, including QNN development, accelerator DSE, and automatic generation of bitstreams and drivers.
It abstracts away the complexity of low-level hardware deployment from deep learning developers and reduces the design effort.
However, FINN's conventional DSE mainly supports heuristic folding exploration for predefined dense QNNs under a targeted throughput, limiting exploration of the model architecture itself.
Complementary efforts improve deployment-side feedback: an empirical quality-of-results (QoR) flow enables faster estimation of dataflow accelerators~\cite{jentzsch2025empirical}, while sparsity has been integrated into its accelerator generation~\cite{li2025logicsparse}.
These works further extend the FINN ecosystem towards more efficient accelerator exploration and deployment.

NAS is increasingly used for design-space exploration of FPGA-accelerated neural networks, where designers can exploit bespoke numeric representations and precisions, in addition to architecture-level optimisations.
NASB~\cite{zhu2020nasb} explores NAS-based architectural optimisation for binary neural networks by searching hardware-suitable binary convolutional cells.
Building on this line of work, NASH~\cite{ji2024nash} extends NAS toward FPGA-oriented QNN deployment and integrates the resulting models into the FINN flow for hardware generation and evaluation.
However, NASH mainly searches network architecture under predefined bit-width settings, while FINN-derived resource and latency costs are evaluated after architecture search rather than incorporated directly into the NAS objective. This leaves scope for search methods that jointly explore architecture and quantisation while using FPGA-oriented cost estimates to guide candidate selection.

To address these gaps, we propose FINNAS, which incorporates proxy validation accuracy and FINN-estimated LUT usage and latency directly into an evolutionary search over quantised MLP topology and precision settings, followed by post-search unstructured pruning (USP).
The main contributions of this paper are listed as follows:

\begin{itemize}
    \item A hardware-aware evolutionary NAS integrated with the end-to-end FINN flow, jointly considering predictive accuracy, estimated LUT usage, and latency during search.
    

    \item An expanded search space over MLP topology and global weight and activation precisions, combined with post-search USP.
    
    \item On the JSC task, FINNAS achieves higher accuracy and lower latency with up to an 8.5$\times$ LUT usage reduction relative to the hand-optimised dense FINN baseline.
\end{itemize}

\begin{figure*}[t]
\centering
\includegraphics[width=0.98\linewidth]{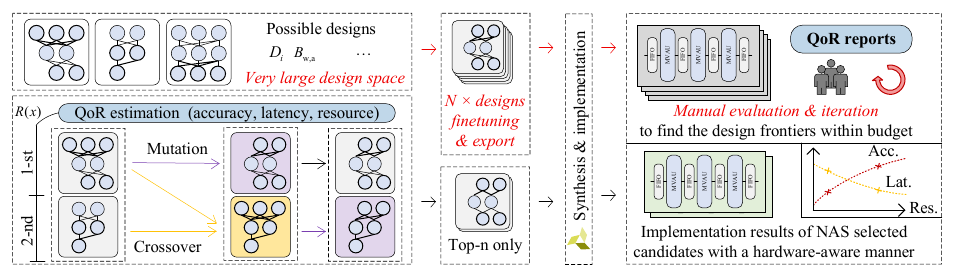}
\caption{FINNAS workflow compared with conventional FPGA design-space exploration.}
    \vspace{-3mm}
\label{fig:nas_overview}
\end{figure*}


\section{Methodology and Design} \label{methodology}

Fig.~\ref{fig:nas_overview} illustrates the overall workflow of the proposed FINNAS framework and compares it with conventional FPGA design-space exploration.
Designing QNN accelerators requires architectural and quantisation choices to be weighted against hardware objectives which, unlike software-only metrics, strongly depend on how a network is mapped to the target FPGA. 
These hardware characteristics are most accurately evaluated through synthesis, but conventional design approaches make this expensive by requiring numerous model variations to be assessed before a suitable trade-off is found.

FINNAS reduces this exploration cost by adopting an evolutionary search algorithm to explore network architectures and quantisation bit-widths, while incorporating FINN-derived hardware estimates into candidate evaluation. This allows us to guide the search towards the most competitive design regions without synthesising every architecture. Selected finalists are subsequently fully retrained, pruned, fine-tuned, and evaluated through RTL simulation and OOC synthesis.

\begin{table}[t]
\centering
\caption{JSC model search space and evolutionary-search settings.}
\label{tab:jsc_search_space}
\small
\setlength{\tabcolsep}{5pt}
\begin{tabular}{lll}
\toprule
\textbf{Parameter} & \textbf{Notation} & \textbf{Setting} \\
\midrule


Hidden depth 
& $L$ 
& $\{2,3,4,5,6\}$ \\

Hidden width 
& $h_i$
& \makecell[l]{$\{8,12,16,24,32,48,64,$\\
$96,128,160,192,256,320\}$} \\


Weight bits 
& $B_w$ 
& $\{2,3,4,8\}$ \\

Input bits 
& $B_{ia}$ 
& $\{3,4,8\}$ \\

Hidden bits 
& $B_{ha}$ 
& $\{3,4\}$ \\

Output bits 
& $B_{oa}$
& $\{3,4,7,8\}$ \\

\midrule

Population size
& $N_{\mathrm{pop}}$
& 30 \\

Generations
& $N_{\mathrm{gen}}$
& 20 \\

Elites per gen.
& $N_{\mathrm{elite}}$ 
& 1 \\ 

Tournament size 
& $k_{\mathrm{tour}}$ 
& 2 \\

Random candidates 
& $N_{\mathrm{rand}}$ 
& 4 \\ 

Crossover prob. 
& $p_c$ 
& 0.3 \\ 

Mutation prob. 
& $p_m$ 
& 0.9 \\ 



\midrule

LUT budget
& $B_{\mathrm{l}}$
& run-dependent \\

Latency budget
& $B_{\mathrm{t}}$
& 15 ns \\

Maximum LUT
& $L_{\mathrm{max}}$
& 250,000 \\

LUT exponent
& $\beta_{\mathrm{l}}$
& 0.50 \\

Latency exponent
& $\beta_{\mathrm{t}}$
& 0.05 \\

\bottomrule
\end{tabular}
    \vspace{-3mm}
\end{table}

\subsection{Hardware-Aware Evolutionary Search} \label{hw_aware_search}

For the exploration of candidate architectures, we adopt an evolutionary algorithm, in contrast to the differentiable NAS strategy used in NASH~\cite{ji2024nash}.
This is particularly suitable for the discrete and constrained architectural and quantisation variables considered here, with candidate fitness obtained from proxy training and FINN hardware estimate reports.
The search parameters used in this work are listed in Table~\ref{tab:jsc_search_space}.

Each individual is encoded as Equation~\ref{eq:0},
i.e., an ordered list of hidden-layer widths and four global precision fields. At initialisation, \(L\), every \(h_i\), and each precision field are drawn from their corresponding finite sets in Table~\ref{tab:jsc_search_space}.
Samples are repaired until they satisfy \(B_{oa}\geq B_{ha}\) and \(B_w<8\) when all activation precisions are below 8 bits.
Each candidate then undergoes 16 epochs of proxy training using cross-entropy loss and the AdamW optimiser to obtain validation accuracy \(A(x)\).
\begin{equation}
    x=\{[h_1,\ldots,h_L],(B_w,B_{ia},B_{ha},B_{oa})\}
    \label{eq:0}
\end{equation}

The candidates are then exported to QONNX and evaluated using the FINN performance estimator to obtain estimated LUT usage \(L(x)\) and latency \(T(x)\), without requiring out-of-context (OOC) synthesis or RTL simulation.
For every candidate, FINNAS applies the same fully parallel mapping by setting \(\mathrm{SIMD}=N_{\mathrm{in}}\) and \(\mathrm{PE}=N_{\mathrm{out}}\) for each fully connected layer; this mapping is retained during final hardware evaluation. 

The estimated accuracy and hardware metrics are combined through the reward \(R(x)\) defined in Equation~\ref{eq:1}.
\begin{equation}
R(x) =
A(x)
\cdot
\max\left(1,\frac{L(x)}{B_{\mathrm{l}}}\right)^{-\beta_{\mathrm{l}}}
\cdot
\max\left(1,\frac{T(x)}{B_{\mathrm{t}}}\right)^{-\beta_{\mathrm{t}}}
\label{eq:1}
\end{equation}
In this equation, \(B_{\mathrm{l}}\) and \(B_{\mathrm{t}}\) denote the LUT and latency budgets, respectively, while \(\beta_{\mathrm{l}}\) and \(\beta_{\mathrm{t}}\) control the penalty strength when an architecture exceeds either budget. 
Candidates satisfying a budget receive no additional reward for further reducing that metric, allowing accuracy to drive selection within the feasible region. 
The algorithm minimises \(F(x)=-R(x)+P(x)\), where \(P(x)=0\) for \(L(x)\leq L_{\mathrm{max}}\) and \(P(x)=1+10(L(x)/L_{\mathrm{max}}-1)^2\) otherwise, ensuring that oversized candidates remain rankable according to their degree of constraint violation.

Within each generation, candidates are ranked according to \(F(x)\) after evaluation; \(N_{\mathrm{elite}}\) elites are retained, and the remaining parents are chosen by tournament selection of size \(k_{\mathrm{tour}}\).
Crossover exchanges hidden-layer suffixes and quantisation fields between parent pairs, while mutation can modify a hidden width, insert or remove a layer, or resample quantisation parameters.
After these operations, a repair step maps each offspring back to the search space by enforcing \(2\leq L\leq6\), restricting every \(h_i\) and bit-width to the sets in Table~\ref{tab:jsc_search_space}, and re-enforcing the two cross-field quantisation constraints used at initialisation.
\(N_{\mathrm{rand}}\) randomly generated candidates are then injected, duplicates are removed, and previously evaluated architectures reuse their cached fitness, overall
increasing exploration and reducing the search cost.
After \(N_{\mathrm{gen}}\) generations, the five highest-ranked architectures in the archive are selected as finalists.

\subsection{Finalist Training and Pruning} \label{pruning}

The selected architectures are retrained with a maximum allowance of 500 epochs and early stopping with a patience of 50 epochs, and first evaluated as dense baselines. The large cap accommodates different convergence rates across the searched topologies, while early stopping limits unnecessary training once validation performance plateaus.
Post-hoc unstructured global magnitude pruning is then applied to their linear layers at different sparsity targets without changing the network tensor shapes.
For each target, a global threshold is determined from the absolute weight values of all eligible weight tensors, with weights below this threshold set to zero.
Subsequently, the pruned networks are fine-tuned for up to 100 epochs when their raw accuracy drops by more than a specified \(\epsilon\), with binary masks enforcing the pruned weights throughout fine-tuning. Test accuracy is recorded on the held-out test set for each dense and pruned finalist.


\section{Experimental Results}\label{results}
\subsection{Experimental Setup}

The model training and accelerator compilation are performed on a workstation fitted with an NVIDIA RTX 5070 Ti GPU and an AMD Ryzen 5 9600X CPU, using CUDA 13.0, Python 3.12 and FINN Docker.
All resource and latency reports are obtained from OOC synthesis and RTL simulation reports in Vivado 2022.2 targeting a Virtex UltraScale+ FPGA (\texttt{xcvu9p-flgb2104-2-i}) with the \texttt{Flow\_PerfOptimized\_high} synthesis strategy. 
The VU9P target matches the device used by most prior JSC implementations considered here; HGQ~\cite{sun2026hgq} instead targets the larger VU13P. 

Across the reported searches, the LUT budget was varied as \(B_{\mathrm{l}}\in\{70,100,150,160\}\)k LUTs to target different points along the accuracy-resource trade-off, while all other search settings were fixed. 
A single NAS run evaluates 500 candidates in around six hours using proxy training and FINN estimates. For context, OOC synthesis of those same 500 sampled candidates would require at least 40 serial hours even at the shortest observed runtime ($\sim$5 min/design).

To validate the search-time hardware proxy, FINN estimates were compared against OOC synthesis and RTL-simulation results for all 20 dense finalists synthesised from the reported searches. Estimated and synthesised LUT usage exhibit a strong Spearman rank correlation of \(\rho=0.854\), while estimated and RTL cycle latency achieve \(\rho=0.781\).
The estimates show substantial absolute error, with LUT usage systematically overestimated and latency represented only crudely, but we found them to preserve sufficient relative ordering to be able to guide candidate ranking during search.




\subsection{Comparison with Prior Work}

\begin{table}[t]
\centering
\caption{Comparison of FINNAS with prior HLF JSC FPGA implementations and FINN baselines}
\label{tab:jsc_cernbox_comparison}
\resizebox{\columnwidth}{!}{%
\begin{tabular}{lccccccc}
\toprule
\textbf{Implementation} 
& \makecell{\textbf{Acc}\\\textbf{(\%)}} 
& \textbf{LUT} 
& \textbf{DSP} 
& \textbf{FF} 
& \makecell{\textbf{Fmax}\\\textbf{(MHz)}} 
& \makecell{\textbf{Latency}\\\textbf{(ns)}} 
& \textbf{II} \\
\midrule
HGQ~\cite{sun2026hgq} 
    & 75.3 & 10,921 & 0 & 11,183 & 578.4 & 31.1 & 1 \\
HGQ~\cite{sun2026hgq} 
    & 75.1 & 5,974 & 0 & 5,775 & 609.8 & 24.6 & 1 \\
AmigoLUT~\cite{weng2025amigolut} 
    & 74.4 & 42,742 & 0 & 4,717 & 520 & 9.6 & 1 \\
AmigoLUT~\cite{weng2025amigolut} 
    & 72.9 & 1,243 & 0 & 1,240 & 1008 & 5.0 & 1 \\
ReducedLUT~\cite{reducedlut2025} 
    & 74.9 & 58,409 & 0 & N/A & 302.8 & -- & -- \\
ReducedLUT~\cite{reducedlut2025} 
    & 72.5 & 2,786 & 0 & N/A & 408.5 & -- & -- \\
NeuraLUT-Asm.~\cite{andronic2025neuralutassemble} 
    & 75.0 & 8,539 & 0 & 1,332 & 352 & 5.7 & 1 \\
NeuraLUT-Asm.~\cite{andronic2025neuralutassemble} 
    & 75.0 & 8,535 & 0 & 2,717 & 994 & 7.0 & 1 \\
QKeras~\cite{coelho2021qkeras} 
    & 74.8 & 39,782 & 124 & 8,128 & $\sim$200 & 55.0 & 1 \\
QKeras~\cite{coelho2021qkeras} 
    & 72.3 & 9,149 & 66 & 1,781 & $\sim$200 & 55.0 & 1 \\
PolyLUT~\cite{andronic2025polylut} 
    & 75.1 & 246,071 & 0 & 12,384 & 203 & 24.6 & 1 \\
PolyLUT-Add~\cite{lou2024polylutadd} 
    & 75.0 & 36,484 & 0 & 1,209 & 315 & 15.9 & 1 \\
NeuraLUT~\cite{andronic2024neuralut} 
    & 75.0 & 92,357 & 0 & 4,885 & 368 & 13.6 & 1 \\
LogicNets~\cite{umuroglu2020logicnets} 
    & 71.8 & 37,931 & 0 & 810 & 427 & 11.7 & 1 \\
\midrule
FINN~\cite{blott2018finn} Dense
    & 73.78 & 57,893 & 0 & 82,750 & 417.54 & 110.17 & 1 \\
FINN~\cite{blott2018finn} 30\%
    & 73.72 & 46,735 & 0 & 67,252 & 438.21 & 104.97 & 1 \\
FINN~\cite{blott2018finn} 50\%
    & 73.63 & 37,386 & 0 & 54,415 & 463.18 & 95.0 & 1 \\
FINN~\cite{blott2018finn} 70\%
    & 73.25 & 10,102 & 0 & 14,684 & 678.43 & 58.96 & 1 \\
\midrule
\textbf{A [32,24] (30\%)}
    & 75.03 & 21,409 & 0 & 10,621 & 451.4 & 71.1 & 1 \\
\textbf{B [16,24,12] (30\%)}
    & 74.78 & 13,829 & 0 & 12,793 & 500.8 & 90.0 & 1 \\
\textbf{C [32,32] (50\%)}
    & 74.53 & 16,233 & 0 & 11,033 & 466.8 & 51.1 & 1 \\
\textbf{D [16,16] (30\%)}
    & 74.54 & 8,125 & 0 & 7,417 & 505.3 & 66.0 & 1 \\
\textbf{E [24,8] (30\%)}
    & 74.36 & 6,806 & 0 & 4,796 & 462.5 & 62.2 & 1 \\
\bottomrule
\end{tabular}%
}
\end{table}



\begin{figure}
    \centering
    \includegraphics[width=0.98\linewidth]{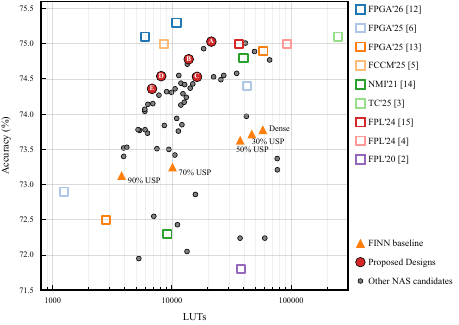}
    \caption{Accuracy-LUT trade-off for HLF JSC FPGA implementations. Selected FINNAS finalists are labelled A-E; additional NAS candidates and pruned FINN baseline variants are also shown.}
    \label{fig:accuracy_vs_lut}
    \vspace{-5mm}
\end{figure}

To evaluate the effectiveness of FINNAS, we conduct a case study on the HLF JSC CERNBox task and compare the searched designs with three groups of baselines: (a) previously reported FPGA accelerators for the same task, (b) a dense FINN implementation using a representative architecture from~\cite{coelho2021qkeras}, and (c) its sparse FINN variants. The prior-work baselines use non-FINN, method-specific accelerator flows, many employing specialised LUT-based mappings, whereas the FINN baseline and Designs A--E are generated through the general FINN dataflow toolchain. The results are summarised in Table~\ref{tab:jsc_cernbox_comparison}.

Designs A--E were selected from searches with \(B_{\mathrm{l}}=\{160,150,100,70,100\}\)k LUTs, respectively, using \(q=(B_w,B_{ia},B_{ha},B_{oa})\), with \(q_A=q_B=(8,8,4,7)\), \(q_C=(4,8,4,7)\), \(q_D=(8,8,4,4)\), and \(q_E=(8,8,3,4)\). Their hidden-layer topologies and selected USP levels are listed in Table~\ref{tab:jsc_cernbox_comparison}.
Among the proposed implementations, Design A achieves an accuracy of 75.03\% using 21.4k LUTs, placing it in the same accuracy range as several specialised LUT-oriented implementations. 
Designs D and E represent the more compact end of the searched design space, maintaining over 74\% accuracy with only 8.1k and 6.8k LUTs, respectively.
Design C achieves the lowest latency among all synthesised designs exceeding 74\% accuracy, while Design B is the highest-accuracy selected design that satisfies the 2\,ns timing constraint.

\begin{table}[t]
\centering
\caption{Dense-to-selected-pruned results for Designs A--E.}
\label{tab:pruning_results}
\small
\setlength{\tabcolsep}{3pt}
\resizebox{\columnwidth}{!}{%
\begin{tabular}{lcccc}
\toprule
& \textbf{Acc. (\%)} & \textbf{LUT} & \textbf{FF} & \makecell{\textbf{Fmax}\\\textbf{(MHz)}} \\
\midrule
A (30\%) & 75.05$\rightarrow$75.03 & 26,667$\rightarrow$21,409 & 14,549$\rightarrow$10,621 & 395.7$\rightarrow$451.4 \\

B (30\%) & 74.84$\rightarrow$74.78 & 16,640$\rightarrow$13,829 & 15,850$\rightarrow$12,793 & 496.5$\rightarrow$500.8 \\

C (50\%) & 74.42$\rightarrow$74.53 & 24,637$\rightarrow$16,233 & 13,009$\rightarrow$11,033 & 461.8$\rightarrow$466.8 \\

D (30\%) & 74.52$\rightarrow$74.54 & 10,360$\rightarrow$8,125 & 9,997$\rightarrow$7,417 & 497.5$\rightarrow$505.3 \\

E (30\%) & 74.48$\rightarrow$74.36 & 8,235$\rightarrow$6,806 & 6,224$\rightarrow$4,796 & 363.9$\rightarrow$462.5 \\
\bottomrule
\end{tabular}%
}
    \vspace{-5mm}
\end{table}

Under the fully parallel mapping used here, USP produces hardware savings because embedded weights become compile-time constants, allowing zero-valued weights to be propagated and corresponding combinational logic removed during synthesis.
In folded mappings, weights instead share time-multiplexed datapaths, so individual zeros do not necessarily remove hardware resources~\cite{blott2018finn}.
Across all synthesised finalists evaluated in the pruning study, 30\%, 50\%, and 70\% USP reduce LUT usage by an average of 14.7\%, 28.6\%, and 39.8\%, respectively, and FF usage by 17.2\%, 26.9\%, and 41.4\%.
This generally comes with a gradual accuracy trade-off for most larger MLPs, while some compact models maintain or slightly improve accuracy.
For the five representative designs A--E, the corresponding dense and selected-pruned results are shown in Table~\ref{tab:pruning_results}.
Pruning also improves Fmax for all five reported designs, enabling timing closure for A, B, D, and E.


Fig.~\ref{fig:accuracy_vs_lut} places the reported implementations in the broader accuracy--LUT design space.
Additional NAS-discovered candidates are also plotted to illustrate the range of trade-offs exposed across the searches.
Relative to the FINN-based designs, FINNAS expands the attainable trade-off by jointly searching network topology and quantisation under hardware-aware resource and latency objectives.
In particular, compared with the dense FINN baseline, Design E reduces LUT usage by \(8.5\times\) and latency by \(1.7\times\), while improving accuracy from 73.78\% to 74.36\%.

Compared with specialised FPGA implementations reported in prior work, FINNAS maintains competitive accuracy at moderate and low LUT usage despite relying on the general-purpose FINN dataflow architecture.
The latency of our designs is not as low as some LUT network implementations that heavily exploit FPGA primitives.
This is mainly because the automated FINN flow preserves the layer-level structure of deep learning models and inserts inter-layer FIFOs to support a more general end-to-end deployment process across models.
Nevertheless, the HLF JSC task is representative of real-time LHC trigger applications, which operate at the 40\,MHz bunch-crossing rate and impose stringent microsecond-scale latency constraints~\cite{cms2020phase2l1trigger}. 
As an end-to-end solution, the FINNAS-generated accelerators therefore provide sufficient latency and throughput margins while achieving a favourable accuracy-resource trade-off under hardware-aware constraints.

\section{Conclusion}\label{conclusion}


In this paper, we present FINNAS, which performs hardware-aware architecture and bit-width search for QNNs, enabling the co-optimisation of accuracy and hardware performance.
On the JSC task, compared with a hand-optimised dense accelerator implemented using FINN, the proposed design reduces LUT usage by up to \(8.5\times\) and latency by \(1.7\times\), while achieving higher accuracy, thereby extending the performance frontier of this end-to-end design framework.
In future work, we will investigate the incorporation of sparse patterns into the search space and introduce inter-layer optimisation into the end-to-end automated design flow to improve accelerator performance.

\bibliographystyle{unsrt}
\bibliography{references}

\end{document}